# Text as Data: Real-time Measurement of Economic Welfare

Rickard Nyman[1] and Paul Ormerod[2]

January 2020

---

[1] Centre for Decision Making Uncertainty, University College London, rickardnyman@gmail.com and Algorithmic Economics Ltd

[2] Department of Computer Science, University College London, p.ormerod@ucl.ac.uk Algorithmic Economics Ltd

## *Abstract*

*Economists are showing increasing interest in the use of text as an input to economic research.*

*Here, we analyse online text to construct a real time metric of welfare. For purposes of description, we call it the Feel Good Factor (FGF). The particular example used to illustrate the concept is confined to data from the London area, but the methodology is readily generalisable to other geographical areas.*

*The FGF illustrates the use of online data to create a measure of welfare which is not based, as GDP is, on value added in a market-oriented economy.*

*There is already a large literature which measures wellbeing/happiness. But this relies on conventional survey approaches, and hence on the stated preferences of respondents.*

*In unstructured online media text, users reveal their emotions in ways analogous to the principle of revealed preference in consumer demand theory.*

*The analysis of online media offers further advantages over conventional survey-based measures of sentiment or well-being. It can be carried out in real time rather than with the lags which are involved in survey approaches. In addition, it is very much cheaper.*

# 1. Introduction

In a recent issue of the *Journal of Economic Literature*, Gentzkow et al. (2019) point out that "New technologies have made available vast quantities of digital text, recording an ever-increasing share of human interaction, communication, and culture. For social scientists, the information encoded in text is a rich complement to the more structured kinds of data traditionally used in research" (p.535).

Here, we provide an empirical illustration of this concept.

We carry out analysis of social media to construct a real time metric of welfare based upon feelings and sentiment, which we call the Feel Good Factor (FGF). The particular example used is confined to data from the London area, but the methodology is readily generalisable to other geographical areas.

The FGF measures in real time the sentiment of the population of Greater London expressed through social media, and in particular on Twitter.

The national accounts were developed in the 1930s in response to the pressing need of policy makers to know what was happening to output. It was very clear that there had been a catastrophic collapse in output in many economies. But a systematic way of measuring output in market-oriented economies had not yet been developed. This was the initial purpose of developing the measurement of GDP.

GDP remains a valuable indicator. But as Jarmin (2019) argues: "The system of economic measurement developed in the 20th century continues to provide critical statistics on the health and performance of the economy. That said, current measurement programs are not keeping pace with the changing economy, and current methods for collecting and disseminating statistical information are not sustainable" (p.180).

Jarmin suggests that "Government statistics in 21st century measurement will be based on vastly more source data, much of which is unstructured—or at least not designed for statistical uses" (p.165).

Series such as this, created by a combination of social media, growth in computing power and developments in machine learning algorithms, are inherently *different* from those in the national accounts.

Such series may or may not prove useful in either understanding or predicting series in the traditional national accounts. This is emphatically not how they should be judged. They represent information which is *additional* to that contained in the national accounts.

Section 2 sets out the basic principles of our approach. Section 3 describes the methodology of obtaining quantitative representations of text before it can be given to machine learning algorithms for classification analysis. In section 4 we consider the choice of machine learning algorithms. Section 5 presents the technical results. Section 6 plots the London Feel Good Factor which is generated and offers a short discussion around this.

## 2. Background

### 2.1 Unstructured text and revealed emotion

An important point to make from the outset is that we are not suggesting that the particular way in which we estimate sentiment in Greater London is completely definitive. As will be apparent, judgement is required in several key places.

However, the conventions which govern the national accounts were not laid down completely at the very beginning of the process. They took time to evolve. Even now, the best part of a century after their initial construction, there is keen debate about how to measure their central concept of real GDP, as the Bean report (2016), for example, evidences.

Our results should be regarded as illustrative of the kind of measurement which can be carried out using information available in text sources. As was the case with the national accounts, consensus will build over time on how to carry out the various steps in the process.

Of course, the sentiment levels of various sectors of the economy are already measured in various ways using conventional survey techniques. Indeed, the Office for National Statistics (ONS) now publishes measures of well-being twice a year, based on a mixture of data such as the unemployment rate and subjective answers to surveys (https://www.ons.gov.uk/peoplepopulationandcommunity/wellbeing/articles/measuringnationalwellbeing/qualityoflifeintheuk2018)

The analysis of text on social media offers three distinct advantages over these conventional approaches.

First, there is a theoretical advantage. Economic theory is built on the principle of revealed preference. Surveys which elicit opinions and answers to hypothetical questions are not as firmly based as the observed actions of agents. Agents reveal their preferences by their decisions. In the same way, in social media conversations, agents reveal their emotions and attitudes.

Second, it can be carried out in real time rather than with the lags which are involved in conventional survey approaches.

Third, it is very much cheaper to construct than measures which are based on conventional survey techniques. Salganik (2019), for example, argues that this type of methodology is typically around 50 times cheaper than standard approaches.

Individual posts may of course contain sarcasm or irony. But the sheer scale of the data which is available suggests that in aggregated measures, any such influences will be swamped by posts which reveal the genuine sentiment of the agent. Even in the limited random sample of tweets available to us in constructing the London Feel Good Factor (described in more detail below), we have over 50,000 tweets each day.

Perhaps more pertinently, huge effort has gone into trying to ensure with conventional survey techniques that the sample reflects the socio-economic composition of the relevant population. This is transparently not the case at the moment with Twitter.

We note that, for all their apparent sophistication, survey techniques do not always achieve their intention. It is well known, for example, that support for right wing parties during election polling is often underestimated, and survey companies devote a lot of effort to try to correct for this. Further, as time goes by, social media publication platforms such as Twitter may well evolve to be more representative of the socio-economic composition of the population.

There is evidence to suggest that emotions and attitudes on Twitter may in practice already be a more or less unbiased indicator of those of the population as a whole. Conventional socio-economic classifications of samples may not be relevant.

For example, we carried out a real time analysis for a commercial client during the Brexit campaign using tweets. The conventional wisdom was that Remain were firm favourites to win. Using standard methods of identifying communities in the social media network (Newman, 2004 and 2006), we, not surprisingly, found that there were two main communities discussing Brexit. The most popular topics in each were quite different. That is, until just over two weeks before the vote itself. The topic of immigration suddenly gained serious traction in what was obviously the Remain community. In contrast, the topics which exercised this community, such as employment, never had any substantial presence in the Leave social media community. It was not possible on this basis to say that Leave would win. But over two weeks before the vote, the chances of a Leave vote were identified to be considerably higher than the conventional wisdom indicated.

We carried out another real time study for a commercial client during the 2017 General Election. This identified within a few days of the election being called that the share of the vote of the two main parties would rise sharply, reversing decades of decline. We showed that Brexit was the principal topic in online conversations, in contrast to perceptions based upon mainstream media. When the British Election Survey published their results in early August 2017, their identification that Brexit had been the most important single topic came as a surprise to many commentators. And during the last few days of the campaign, social media analysis indicated a further slight shift away from the Conservatives, in contrast to most of the survey based conventional polls[3].

## 2.2 The data base

Our measurement of sentiment in Greater London (the Feel Good Factor, FGF) is based on the 1 per cent random sample of tweets which is provided for free by Twitter[4].  We used the Java library Twitter4J[5] to access the official Twitter API.   Identifying those located within the Greater London area gives us 66 million tweets in total since we began the analysis on 15 June 2016.

These are tweets about *any* topic, football, holidays, going to the pub, your job, your commute to work, whatever.  No screening of content is carried out.  We take the raw material of the tweets.  This is by design, as our intention is to create a general indication of wellbeing as a whole. One can readily focus the measure on particular topics with a more specific selection of tweets.

The field of sentiment analysis of text data is moving rapidly. A few years ago, a popular way of doing this was based on a count of specific words whose emotional content had been established by surveys or experimental work separate to the text being studied. An example is Associative Norms for English Words (Bradley and Lang, 1999).  A similar approach, but which does not even use word lists validated in psychological experiments, is adopted by Baker et al. (2016).

This approach has now been overtaken in machine learning analysis. Machine learning algorithms are classifiers. So, too, is logistic regression, to give an example familiar to economists.  In order to be able to classify the emotional content of a tweet as either positive or negative, we need to present the algorithms with a set of tweets which labels them as being either positive or negative.

The algorithm does not simply count specific words from a pre-assigned list whose emotional content is measured outside the text. It learns the emotion directly from the text, as we describe in detail below.

There are several potential ways in which such a set could be developed.  Here, we rely on the emojis which are contained in many tweets.  In Greater London, around 20 per cent of tweets contain emojis.  Over the full data period, 15 June 2016 to the end of 2019, this gives us some 13 million tweets with emojis.

---

[3]   More   detailed   information   on   both   these   studies   is   available   at http://www.algorithmiceconomics.com/applied/

[4] See https://dev.twitter.com/streaming/reference/get/statuses/sample for details on the API

[5] See http://twitter4j.org/en/ for details on the Java library used to access the Twitter API

For training the classifier, we choose a random sample from this set of 10,000 positive tweets and a random sample of 10,000 negative tweets. We experimented with different sample sizes, but once the sample size for each category is more than just a few thousand, the results seem very robust with respect to sample size.

To label positive tweets we select tweets containing any of:

'😊|🙂|😊|😃|😁|😀|😄|😆'

But not containing any of:

'😂|☹️|😠|😡|😢|😭|😣|😩'

To label negative tweets we select tweets containing any of:

'☹️|😠|😡|😢|😭|😣|😩',

But not containing any of:

'😂|😊|🙂|😊|😃|😁|😀|😄|😆'

We do not use any tweets containing the emoji 😂. Clustering analysis of emojis indicates that this is used quite independently of others. Inspection of a small sample of tweets containing this emoji indicated it is often used in both a positive and negative context.

Once we have selected the tweets, we *exclude* from them the emojis used for labelling. By doing so we force the machine learning classifiers to only look at the text of the tweets when deciding on their emotional content.

Of course, a slightly different set of emojis could be used. There are many other emojis indicating a variety of emotions and there are new ones regularly being created by tech companies such as Apple. Emojis of various kinds are becoming increasingly important means of communication on online media. The ones we use here seem to us a good starting point.

**2.3    A benchmark**

We ran a small experiment to try to establish a benchmark as to how accurate a machine learning analysis of such tweets might be. We selected 50 positive and 50 negative tweets at random from the populations of positive and negative London tweets, using emojis to determine whether they were positive or negative.

We then shuffled the tweets and put them into an Excel file. We again stripped out all the emojis, leaving just the text of the tweets.

We sent the file to 11 employees of a consultancy company based in London and asked them to fill in whether the tweets were positive or negative, based purely on the text, by entering a 1 or a 0 against each tweet.

All the people involved work in London and so are familiar with the social and cultural environment of the capital. All except one is trained as either an economist or a psychologist. Most of them are in their 20s, and so have a natural familiarity with online media in general. We might therefore reasonably expect this small group to be better than a random selection from the UK population at judging the emotional content of a tweet.

Further, the tweets they were given are ones which have emojis attached to them and so might reasonably be assumed to have definite emotional content. A random selection of tweets from the entire population, rather than the subset of this which contain emojis, would be harder to classify.

We summed the choices made by the 11 humans across the 100 tweets. We assume that if a tweet scores 0, 1 or 2, it is definitely negative, and if it scores 9, 10 or 11 it is definitely positive. There are 82 of these in total. The remaining 18 score between 3 and 8, essentially distributed equally across these scores. In other words, in 18 per cent of the tweets, humans do not agree as to whether the emotional content is positive or negative.

There is some suggestion that the humans rank the collection of tweets more positively than the emojis. But on a standard chi-square test, the null hypothesis that the human distribution is 50/50 between positive and negative is only rejected at a p-value of 0.101.

However, the main point is that even with a group of humans who might reasonably be expected to be considerably more expert at classifying tweets than a group chosen at random from the UK population, and even when they classify tweets which the senders intend to have emotional content (from the fact that they attach positive or negative emojis to them), they achieve an accuracy of 82 per cent. This suggests there is an upper limit as to what might be expected to be achieved in terms of classification accuracy. The emojis themselves are of course attached to tweets by humans, and so there will be some inherent uncertainty about the labelling.

### 3. Converting text to data

In this section we discuss the conversion of the text of the tweets into some form of quantitative representation for analysis by machine learning algorithms. We use methods which are standard in machine learning.

The approach (known as GloVe) described by Pennington et al. (2014) is widely used and the paper has over 6,000 citations. A clear overview, with a description of how to download and use the method, is given at https://nlp.stanford.edu/projects/glove/.

The authors assemble a very large corpus of words from sources such as Twitter, Wikipedia and a site which crawls web pages. A co-occurrence matrix is constructed, which describes how frequently pairs of words co-occur with each other in any given corpus.

The referenced webpage above states: "The training objective of GloVe is to learn word vectors such that their dot product equals the logarithm of the word's probability of co-occurrence. Owing to the fact that the logarithm of a ratio equals the difference of logarithms, this objective associates (the logarithm of) ratios of co-occurrence probabilities with vector differences in the word vector space".

The eventual output of the process is that every word in the corpus has a unique n-dimensional vector associated with it. The elements of each vector are real valued numbers which essentially describe the closeness of the word to all other words in the corpus. This description is perforce rather imprecise, and is only intended to give a broad non-technical indication of what is going on. As noted above, full technical details are in Pennington et al. (op.cit.).

A tweet with k words will therefore have k vectors associated with it (if a word appears twice, say, the same vector will appear twice). We average these vectors to generate a vector associated with each tweet.

The machine learning algorithms carry out pattern recognition of these average vectors in order to classify the (labelled) tweets into the positive or negative categories.

In terms of the dimension of the vectors, n, the higher is n, the greater are the potential dissimilarities between individual words. In the limit, for example, we might imagine that each vector is of dimension M, where M is the total number of words in the corpus. Each word has a vector containing M-1 zeros and a single 1. But these vectors would be of no practical use. They form an orthogonal space. So, for example, a close synonym of a given word would not be recognised as such.

The GloVe approach clearly involves a very substantial amount of dimension reduction. We can usefully think of there being a trade-off between capturing the differences between words accurately, and being able to recognise synonyms of a word. An obvious example would be the words "mobile" and "cell" before the word "phone". At one level, these words are completely different. But they mean the same thing in this particular context. Ideally, we want to compress the differences as much as possible whilst at the same time preserving the relationships between words which enable a machine learning classifier to distinguish between the positive and negative tweets.

We also examine a variant of the GloVe methodology. The corpus of English words used by GloVe may of course be dominated by sources from the United States and more generally by American English.

Our purpose is to construct a real-time indicator of sentiment tweets originating in Greater London. Most of these will obviously use British English. Further, their content, whilst very wide ranging as any casual inspection of individual tweets chosen at random will show, will tend to reflect the preoccupations of Londoners rather than more global ones. Of course, many tourists in London will tweet, but the resident population of London is now almost 9 million.

Formally, we make use of the word2vec methodology described in Mikolov et al. 2013. The paper, at the time of writing, has almost 9,000 citations.

The basic idea is as follows. Suppose we have a sequence of n words, where n is an odd number. We leave out the middle word and try and predict what it is from the surrounding ones. In a large corpus of text, any given middle word (for example "phone") may appear many times. It will be surrounded on different occasions by different sets of words in different sequences. The neural network attaches weights to all of these words when it is trained to predict the word "phone". These weight vectors are the word vectors we use

Mikolov et al. show that the relatively simple neural network architectures they propose achieve high quality word-vector representations comparable to more complex networks structures such as Recurrent Neural Nets while being much faster to train on larger data sets. Two main architectures were proposed and we make use of the variant they call Continuous Bag-of-Words (CBOW). We use the same n as in GloVe, which is the 5 words both immediately preceding and following any given word.

Our training set for this is 10 million tweets chosen at random from the corpus of the over 60 million tweets in Greater London which we have. (It may perhaps already be apparent that online media data merits the adjective "big". Sample sizes can be obtained readily which dwarf almost all conventional data sets used by economists)

## 4. Choice of ML algorithm

In this section we consider which algorithms to use in order to classify the tweets into positive and negative categories.

Guidance is offered by Fernandez-Delgado et al. (2014), in a paper whose citations are rising rapidly. The authors compare 179 classification algorithms from 17 "families" such as Bayesian, neural networks, logistic and multinomial regression. They examine their performance on 121 data sets in the University of California at Irvine machine learning repository. This repository is in standard use in machine learning research. The authors find that the random forest family of algorithms achieves the best results. The closest rival is support vector machines. There are a few of others which have good results. But the authors note that the remainder, which include Bayesian and logistic regression algorithms, "are not competitive at all". (p.3175).

We therefore examine the performance of random forest (RF) and support vector machines (SVM) algorithms. We examine both linear and radial basis function kernel SVMs (RBF).

Random forests (Breiman, 2001, 2002) are machine-learning models known for their ability to cope with noisy, non-linear, high-dimensional prediction problems. Many proofs of their properties which extend the original work of Breiman are available in, for example, Biau et al. (2008) and Biau (2012).

They construct a large number of decision trees in training by sampling with replacement from the observations. Each tree in the collection is formed by first selecting at random, at each node, a small group of input coordinates to split on and, secondly, by calculating the best split based on these features in the training set. Each tree gives a prediction, and the predictions are averaged. From the point of view of the bias-variance trade-off, the ensemble of a large number of trees trained on independent bootstrap samples, each with relatively large variance but low bias, achieve much reduced variance without the introduction of additional bias.

More formally, there are a few different variants of random forest classifiers, each built with varying levels of 'randomness'. All variants construct decision trees from samples of the data. A decision tree can be considered a sequence of logical rules applied to a selection of features that ultimately, in the case of classification, assigns an observation to one of the possible classes.

A Random Forest algorithm in its most common form constructs a number of decision trees through a form of bootstrap aggregating (*bagging)*. Several trees are fitted on random samples of the training set $X = x_1, \ldots, x_N$ with corresponding class labels $Y = y_1, \ldots, y_N$. To minimize the correlation between trees, the data is often sampled with replacement in order for each tree to observe a slightly different set of observations.

Formally, for $b = 1, \ldots, B$ (where *B* is the number of trees to be built):

1. Sample, with replacement, *N* training samples from *X, Y*. Call the samples $X_b, Y_b$
2. Train a classification tree $f_b$ on $X_b, Y_b$

Furthermore, when training each tree $f_b$ a random subset of the features in $X_b$ is considered for each split to ensure that all trees do not use the most predictive features of the training data during construction, to further decrease the correlation between the trees.

When classifying unseen instances each tree makes a classification, and the class mode is selected as the final classification. In the case of regression, the average can be used,

$$\mathcal{F}(x) = \frac{1}{B} \sum_{b=1}^{B} f_b(x)$$

An estimate of the uncertainty of each prediction can be made as the standard deviation of all predictions,

$$\sigma = \sqrt{\frac{\sum_{b=1}^{B} f_b(x) - \mathcal{F}(x)}{B - 1}}$$

An estimate of the confidence of a particular classification can similarly be made using the proportion of trees classifying the observation as the mode class.

A classic reference for support vector machines is Cortes, C. and Vapnik, V. (2013), which has over 30,000 citations. We are carrying out a binary classification (is the tweet positive or negative?) with N data points of dimension p. SVMs find the p-1 dimensional hyperplane which optimally separates the two classes. In other words, it is trying to maximise a distance metric between them. The data may not of course be linearly separable, and the linear SVM algorithm contains a hyperparameter whose value determines the trade-off between the distance and the number of correctly classified examples. The radial basis variant (RBF) maps the original data points into a higher dimensional space in which it is more likely to find a plane which separates the two classes completely.

More formally, SVMs find a function,

$$f(x) = w\mathbf{x} + b,$$

Which minimises the norm of the hyperplane $w$, $\|w\|^2$, while satisfying the constraints,

$$wx_i - b \geq 1, y_i = 1$$
$$wx_i - b \leq -1, y_i = -1, \forall i$$

This can be equivalently written as "minimise $\|w\|^2$ subject to $y_i(wx_i - b) \geq 1, \forall i$"

Even if the data are linearly separable, strictly forcing the classifier to assign each point to the correct side of the separating plane might lead to poor generalisation. The constraint can be relaxed by introducing a loss function (*hinge* loss),

$$\max(0, 1 - y_i(wx_i - b))$$

which is zero for each observation $x_i$ that satisfies the original constraint, but which is proportional to the distance between the plane and each point not satisfying the constraint.

We may now instead minimise the following function

$$\left[\frac{1}{N} \sum_{i=1}^{n} \max(0, 1 - y_i(wx_i - b))\right] + C\|w\|^2$$

The parameter *C* determines the trade-off between increasing the margin size between the separating plane and the nearest points of each class and ensuring data points are on the correct side of the plane. For small values of *C*, the second term will lose importance and the SVM will attempt to separate each class perfectly but possibly with a small margin and for larger values of *C*, the SVM might ignore some points in order to achieve a wider separating margin between the two classes.

If the data are not linearly separable, the input feature space can be transformed by a so-called *kernel function*. In the case of the RBF kernel, we transform each input vector $x_i$ to a vector with features,

$$\exp\left(-\gamma \|x_i - x_j\|^2\right),$$

for a set of reference points $x_j$. Here, $\gamma$ is a further hyperparameter controlling the 'similarity' between $x_i$ and the reference points $x_j$. Thus, we can greatly increase the dimensionality of the original input space and ensure the data are linearly separable in the new representation.

## 5. Results

The machine learning community places great emphasis on what econometricians refer to as out-of-sample performance. As mentioned in section 2, we train the algorithms on 10,000 tweets selected at random from those containing at least one positive emoji, and 10,000 selected at random from those containing at least one negative emoji.

We use the approach of 10-fold cross validation in training the algorithms. We partition the data into 10 "folds" of equal size. An algorithm is trained using the first 90 per cent of the data, and the first "fold" containing the remaining 10 per cent of the data is then predicted. The second fold is then predicted, after training the model on the second set of 90 per cent of the entire data, and so on.

The results set out below are therefore all based upon the out-of-sample performance of the models.

Once we have selected a model using this methodology – and only then – do we present the model with a completely new, previously unseen, set of 10,000 positive tweets and 10,000 negative tweets, again selected at random from the set of tweets containing the relevant emojis but which were not used in the training process. Slightly confusingly, given that we are using 10-fold cross validation in the training process, these new sets of data are referred to by the machine learning community as being the "validation" process.

We stress that this is exactly the procedure we followed. The applied econometric community, certainly in time series analysis, also places emphasis by out-of-sample performance. However, the practice of re-specifying the model if it initially fails out of sample stability tests is widespread amongst econometricians. This was emphatically not done here.

There is a potentially very large grid of options to search in terms of which model to use in the validation step described in the above paragraph. We need to examine: which algorithm to choose; which hyperparameters of the chosen algorithm to select; whether to use the GloVe or Word2Vec word vector representation; and, finally, the dimension of the word vectors.

Initially, we examined the relative performance of the GloVe and Word2Vec approaches, and investigated the appropriate dimension of the word vectors to use in further analysis.

We compare the performance of the random forest, linear SVM and radial basis SVM using the default values of the hyperparameters in Python. We use the algorithms available in the Scikit-Learn Python package. More specifically, the *LinearSVC* function to train and evaluate the linear SVM, the *SVC* function for the SVM with RBF kernel and the *RandomForestClassifier* function to train and evaluate random forests.

We report the percentage of tweets which are correctly classified taking the average of the performance out-of-sample in each of the 10 folds, and the variance of this number across the 10 folds.

The results of the 10-fold cross validation are set out below.

**Table 1.    10-fold cross validation of 20,000 tweets**

| Word Vectors | Dimensionality | Model | Mean Accuracy | SD Accuracy |
|---:|---:|---:|---:|---:|
| GLOVE | 25 | Random Forest | 0.72980 | 0.011645 |
| GLOVE | 25 | Linear SVM | 0.72065 | 0.009050 |
| GLOVE | 25 | RBF SVM | 0.72995 | 0.011197 |
| GLOVE | 50 | Random Forest | 0.74550 | 0.010555 |
| GLOVE | 50 | Linear SVM | 0.75355 | 0.007019 |
| GLOVE | 50 | RBF SVM | 0.75535 | 0.008658 |
| GLOVE | 100 | Random Forest | 0.74935 | 0.009953 |
| GLOVE | 100 | Linear SVM | 0.76480 | 0.009001 |
| GLOVE | 100 | RBF SVM | 0.76400 | 0.008465 |
| GLOVE | 200 | Random Forest | 0.75065 | 0.011404 |
| GLOVE | 200 | Linear SVM | 0.77535 | 0.009053 |
| GLOVE | 200 | RBF SVM | 0.76615 | 0.009587 |
| Word2Vec | 25 | Random Forest | 0.75765 | 0.010861 |
| Word2Vec | 25 | Linear SVM | 0.75670 | 0.009448 |
| Word2Vec | 25 | RBF SVM | 0.77165 | 0.007039 |
| Word2Vec | 50 | Random Forest | 0.76375 | 0.008112 |
| Word2Vec | 50 | Linear SVM | 0.77175 | 0.007481 |
| Word2Vec | 50 | RBF SVM | 0.78440 | 0.007987 |
| Word2Vec | 100 | Random Forest | 0.76420 | 0.013096 |
| Word2Vec | 100 | Linear SVM | 0.78425 | 0.010861 |
| Word2Vec | 100 | RBF SVM | 0.79505 | 0.008498 |
| Word2Vec | 200 | Random Forest | 0.76845 | 0.009799 |
| Word2Vec | 200 | Linear SVM | 0.79220 | 0.010448 |
| Word2Vec | 200 | RBF SVM | 0.79860 | 0.009412 |
| Word2Vec | 350 | Random Forest | 0.76590 | 0.009033 |
| Word2Vec | 350 | Linear SVM | 0.79415 | 0.009047 |
| Word2Vec | 350 | RBF SVM | 0.79855 | 0.008347 |
| Word2Vec | 500 | Random Forest | 0.76585 | 0.009563 |
| Word2Vec | 500 | Linear SVM | 0.79530 | 0.009498 |
| Word2Vec | 500 | RBF SVM | 0.79735 | 0.007517 |

The mean accuracy is the percentage of tweets which are correctly classified, averaging the results of each fold of the 10-fold process. SD accuracy is the variance of the accuracy across the 10-folds.

The dimensionalities for the GloVe approach are those which are made available on the GloVe website. We are able to construct higher dimensional vectors for the Word2Vec approach. However, there is no improvement beyond 200. The results for both GloVe and Word2Vec for dimensions 100 and 200 are very similar. They are clearly superior to dimension 25, and in general better than dimension 50.

Overall, the Word2Vec classifications are slightly more accurate for any given algorithm and dimension than the GloVe classifications. This is not unexpected. Although the GloVe vectors are obtained from a much larger corpus, the Word2Vec vectors are generated from content which is specific to the Greater London area.

Our next step was to examine the effect of different values of the hyperparameters of the algorithms, applying them to the Word2Vec data set with 200 dimensions.

The first step was to examine the results using random forests and linear SVMs. These results are set out in Tables 2a and 2b.

**Table 2a. 10-fold cross-validation, random forest, Word2Vec 200-dimensional word vectors, different values of the hyperparameters**

| Mean Accuracy | SD Accuracy | n_estimators | bootstrap | max_depth |
|---|---|---|---|---|
| 0.77320 | 0.012663 | 100 | True | NaN |
| 0.77935 | 0.011064 | 300 | True | NaN |
| 0.74125 | 0.011518 | 100 | True | 3.0 |
| 0.74265 | 0.010750 | 300 | True | 3.0 |
| 0.77760 | 0.012140 | 100 | False | NaN |
| 0.78145 | 0.011988 | 300 | False | NaN |
| 0.73945 | 0.010622 | 100 | False | 3.0 |
| 0.74065 | 0.010840 | 300 | False | 3.0 |

The column "n_estimators" shows the number of trees in the forest. "Bootstrap" indicates whether in the generation of each tree the sampling is done with or without replacement. "Max_depth" shows the maximum depth of each, and NaN indicates no limit to the splits.

These results suggest that the choice of sampling makes little difference, but that more trees and more depth improve classification accuracy.

**Table 2b. 10-fold cross-validation, linear SVM, Word2Vec 200-dimensional word vectors, different values of the hyperparameters**

| Mean Accuracy | SD Accuracy | C |
|---|---|---|
| 0.79830 | 0.011167 | 0.1 |
| 0.79840 | 0.010809 | 0.5 |
| 0.79715 | 0.012178 | 1.0 |
| 0.78145 | 0.011284 | 10.0 |
| 0.73270 | 0.046915 | 20.0 |
| 0.71100 | 0.035847 | 30.0 |
| 0.68810 | 0.045209 | 40.0 |
| 0.67330 | 0.055693 | 50.0 |

The hyperparameter "C", as described in section 4 above, determines the trade-off between increasing the margin size between the separating plane and the nearest points of each class and ensuring data points are on the correct side of the plane. It is clear that low values perform better.

Before choosing between random forests and SVMs, we examined whether the RBF non-linear performed better, searching extensively over values of the second parameter, γ. Full details of the results are available on request. But no advantage was conferred.

The results obtained with the linear SVM approach do seem slightly better than those of the random forest.

Table 3 below sets out what econometricians refer to as the "contingency table" and the machine learning community describes as the "confusion matrix" for the random forest and SVM with the highest classification accuracy in Tables 2a and b respectively.

**Table 3.** Contingency tables ("confusion matrix") for linear SVM with C = 0.5 and random forest with bootstrap = false, max.depth = unlimited and n_estimators = 300

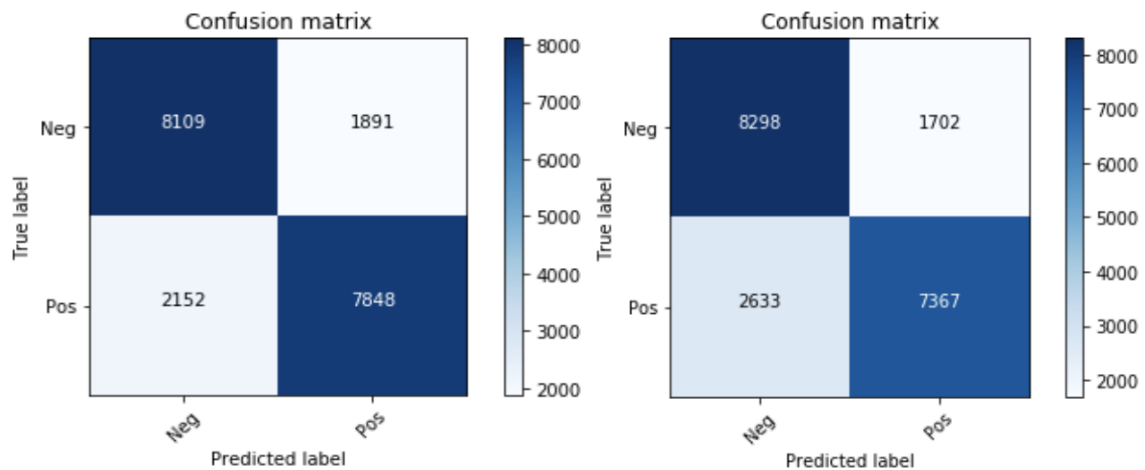

The table on the left is the SVM and on the right the random forest. The SVM gives a slightly higher number of correct classifications. Further, the errors are more balanced, with the random forest errors being weighted towards classifying positive tweets as negative.

We therefore choose the SVM model for the validation stage. Table 4 below shows the confusion matrix when it is presented with an entirely new set of 10,000 positive and 10,000 negative tweets.

**Table 4.** Confusion matrix for the validation process, results on 10,000 positive and 10,000 negative previously unseen tweets

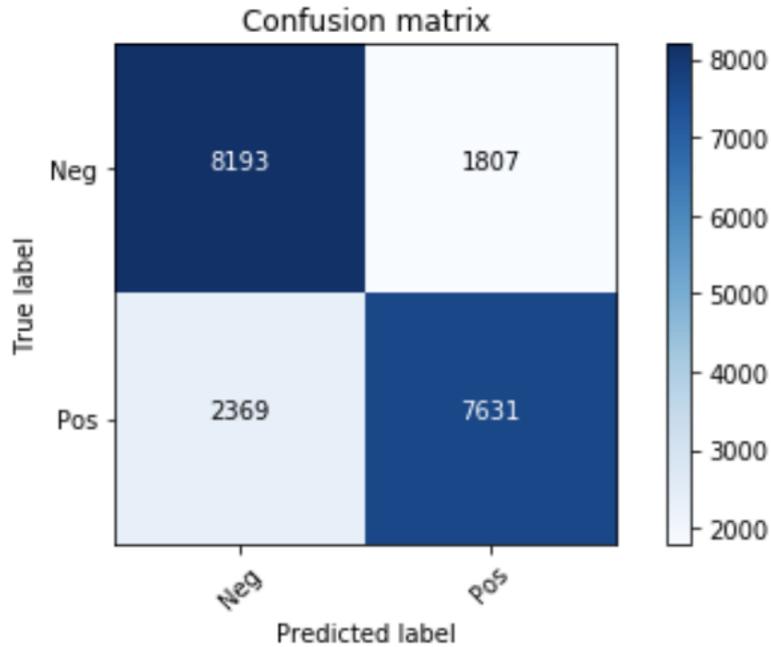

In the training process, 79.8 per cent of tweets are classified correctly, and 79.1 per cent in the previously unseen set of 20,000 tweets.

Most of the classification errors occur with tweets which are reasonably close to being categorised in their correct class. For example, of the true negatives which are actually classed as positive, only 3 per cent are assigned a probability of being positive which is greater than 0.9. 34 per cent are assigned a probability of being positive of between 0.5 and 0.6, and a further 25 per cent between 0.6 and 0.7.

## 6. The London Feel Good Factor

The Feel Good Factor which we obtain by the above process is plotted in Figure 1 below.

**Figure 1.  London Feel Good Factor, 15 June 2016 to 2 January 2020**

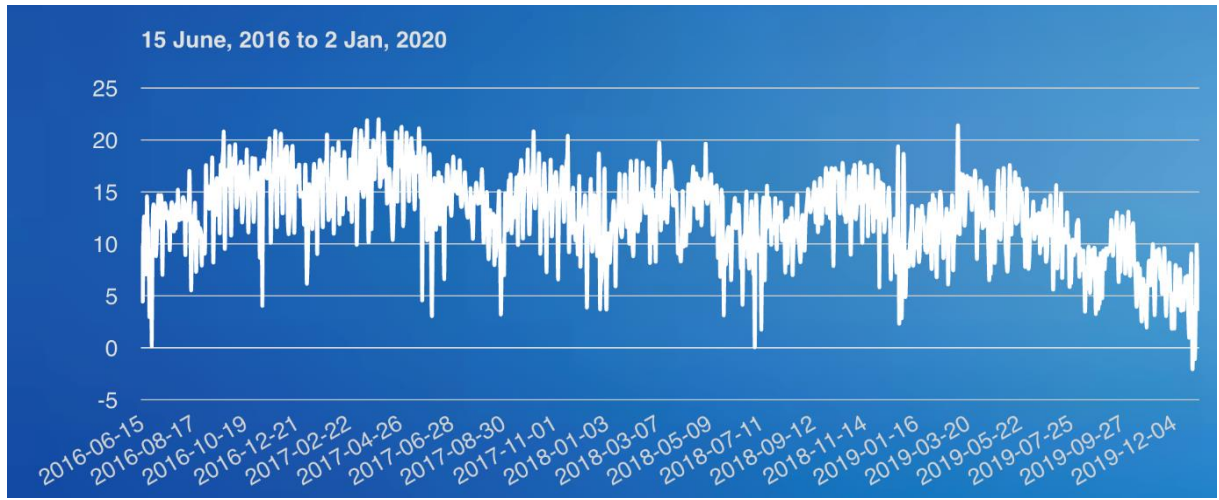

There is very clear evidence of a day of the week effect.  Figure 2 below plots the smoothed power spectrum of the series.  We use the command "spec.pgram" in R.  This calculates the periodogram using a fast Fourier transform, and optionally smooths the result with a series of modified Daniell smoothers.  The raw periodogram is not a consistent estimator of the spectral density, but adjacent values are asymptotically independent. Hence a consistent estimator can be derived by smoothing the raw periodogram.

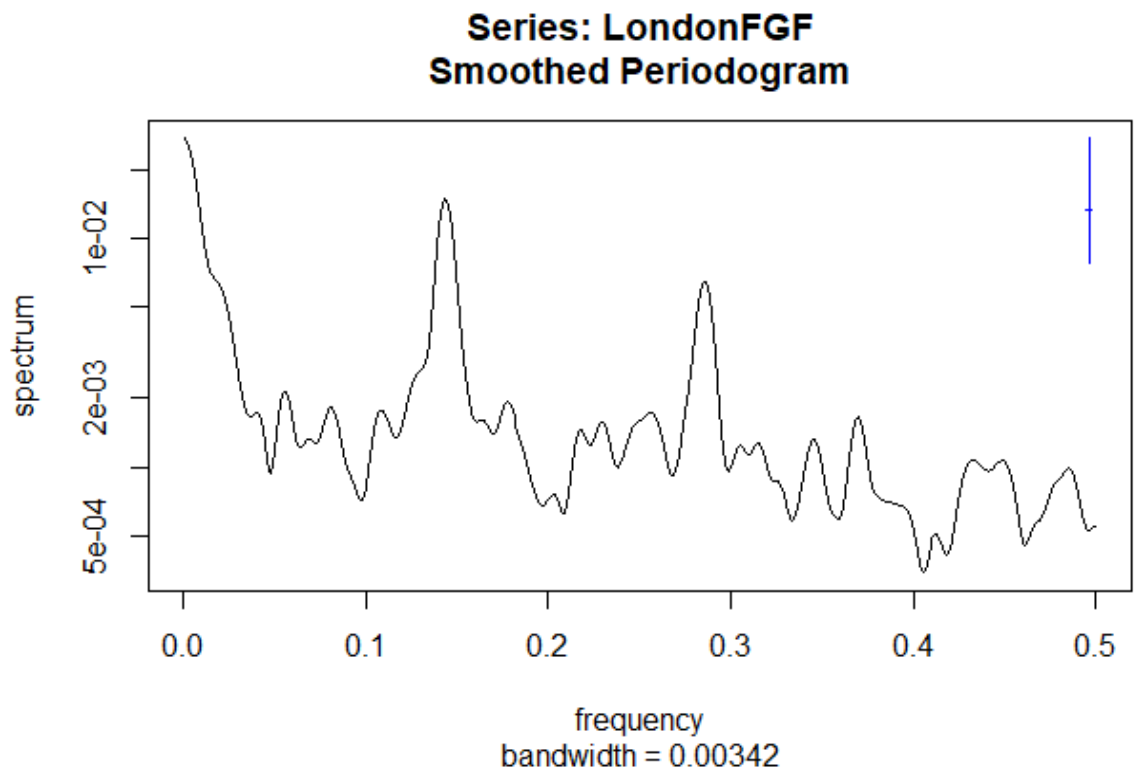

There is a distinct peak at frequency values around 0.14, which of course corresponds to 7 days. There is a second peak to the right of this around 0.28, suggesting that the week can be split into two in terms of sentiment. A simple regression of the series on day of the week dummies certainly suggests that sentiment is consistently higher on Thursday, Friday and Saturday than it is on Sunday and Monday.

The power spectrum also takes high values at very low frequencies, suggesting there is some kind of quarterly or half-yearly seasonality in the data, though given that we only have two and a half years of data we cannot be definitive about its exact nature.

The series clearly varies considerably, but it passes several tests of common sense. For example, in each year at Christmas and New Year, there are substantial fluctuations. The series rises on 24 and 25 December, falling back sharply on 26 December. There are further peaks on 31 December and 1 January, before dropping on 2 and 3 January.

The first low point (on the very far left of the chart) is 24 June 2016. This was of course the day immediately after the Brexit referendum, when London voted Remain and the rest of England voted Leave. This was a Friday, and there is some recovery in the series over the weekend. But Monday 27 June shows another very low reading, presumably as people went back to work and moaned about the result with their colleagues.

There is a sharp low point on 9 November 2016, the day that liberal London learned that it was to be President Trump rather than President Clinton.

There are several more general observations which are worth making about the series. For example, the majority of the economics profession in the UK do seem to have genuinely believed that a Leave vote would precipitate an immediate economic recession. The Treasury forecast dubbed "Project Fear" are just one manifestation of this view.

However, in real time, it was apparent from the London FGF that there was no evidence of a fall in the overall level of sentiment of Londoners. If anything, it rose during the rest of 2016.

The second is that, whilst there are obviously fluctuations, the series does peak in the spring/early summer of 2017 and there is a slight downward trend since then. This corresponds to the slowing down and fall, for example, in London house prices.

Finally, since the spring of 2019 there is an obvious negative trend in the series, reflecting the gloom of Remain London when confronted with the imminence of Brexit.

## 7. Concluding remarks

We illustrate here how to combine machine learning techniques with unstructured online media text databases to create a data series which is the first step in fulfilling Jarmin's prediction that "Government statistics in 21st century measurement will be based on vastly more source data, much of which is unstructured" (op. cit.)

As we stressed early in the paper, the analysis is not meant to be definitive. It does, however, illustrate the kinds of series which can now be created using a combination of "Big Data" (more specifically, online media), and advances in both computing power and machine learning algorithms.

We suggest that the Feel Good Factor has advantages over the more conventional measures of wellbeing/happiness which have been developed. The latter depend upon survey approaches and hence upon the stated preferences of respondents. The FGF is based upon emotions revealed by users in unstructured text data. The FGF is also not only much cheaper to construct, but it is available in real time.

We are not of course suggesting that the series is an infallible guide to the economic prospects of London. It does, however, provide policy makers with real time information which previously simply could not have been generated. For example, the UK Treasury had predicted an immediate recession in the second half of 2016 if Britain voted Leave in the Brexit referendum of June of that year. The FGF showed that there was no sign of a collapse in mood in London. Indeed, the population seemed to become more content after the Brexit vote.

It is possible, though this is speculation on our part, that politicians may find series such as this more useful than the crown jewel of the nation accounts, real GDP. Movements in output of course affect the economic welfare of the electorate. But the feelings and sentiment of the electorate may be a much more tangible series for elected politicians.